\documentclass[journal]{IEEEtran}
\usepackage[T1]{fontenc}

\ifCLASSINFOpdf
\else
\fi

\usepackage{latexsym}
\usepackage{amsmath,amssymb,enumerate}
\usepackage{amsthm}
\usepackage{bm}
\usepackage{setspace}
\usepackage{cite}
\usepackage[pdftex]{color}
\usepackage[pdftex]{graphicx}
\usepackage{comment}
\usepackage{multirow}
\usepackage{scalefnt}
\usepackage{booktabs}
\usepackage{multirow}
\usepackage{algorithm,algorithmic}
\usepackage{mathrsfs}

\newcommand{\diag}{\mathrm{diag}}

\newcommand{\TT}{\top}

\newcommand{\LL}{\bm{\mathcal{L}}}
\newcommand{\Lrw}{\bm{\mathcal{L}}_{\mathrm{rw}}}
\newcommand{\Laug}{\bm{\mathcal{L}}_{\mathrm{aug}}}
\newcommand{\Lmax}{\bm{\mathcal{L}}_{\max}}

\begin{document}
%
\title{Interpreting Graph-based Sybil Detection Methods as Low-Pass Filtering}
%
%
%

\author{Satoshi~Furutani,         
        Toshiki~Shibahara,
        Mitsuaki~Akiyama,~\IEEEmembership{Member,~IEEE,}
        and~Masaki~Aida,~\IEEEmembership{Senior Member,~IEEE}
\thanks{Manuscript received June xx, 2022; revised August xx, 2022.}%
\thanks{
S. Furutani, T. Shibahara, and M. Akiyama are with NTT Social Informatics Laboratories, Tokyo 180-8585, Japan (e-mail: satoshi.furutani.ek@hco.ntt.co.jp; toshiki.shibahara.de@hco.ntt.co.jp; akiyama@ieee.org).}
\thanks{
S. Furutani and M. Aida are with Graduate School of Systems Design, Tokyo Metropolitan University, Tokyo 191-0065, Japan (e-mail: furutani-satoshi1@ed.tmu.ac.jp; aida@tmu.ac.jp).}}

%
%

\markboth{Journal of \LaTeX\ Class Files,~Vol.~14, No.~8, August~2015}%
{Shell \MakeLowercase{\textit{et al.}}: Bare Demo of IEEEtran.cls for IEEE Communications Society Journals}
%



\maketitle

\begin{abstract}
Online social networks (OSNs) are threatened by Sybil attacks, which create fake accounts (also called Sybils) on OSNs and use them for various malicious activities.
Therefore, Sybil detection is a fundamental task for OSN security.
Most existing Sybil detection methods are based on the graph structure of OSNs, and various methods have been proposed recently.
However, although almost all methods have been compared experimentally in terms of detection performance and noise robustness, theoretical understanding of them is still lacking.
In this study, we show that existing graph-based Sybil detection methods can be interpreted in a unified framework of low-pass filtering.
This framework enables us to theoretically compare and analyze each method from two perspectives: filter kernel properties and the spectrum of shift matrices.
Our analysis reveals that the detection performance of each method depends on how well low-pass filtering can extract low frequency components and remove noisy high frequency components.
Furthermore, on the basis of the analysis, we propose a novel Sybil detection method called SybilHeat.
Numerical experiments on synthetic graphs and real social networks demonstrate that SybilHeat performs consistently well on graphs with various structural properties.
This study lays a theoretical foundation for graph-based Sybil detection and leads to a better understanding of Sybil detection methods.
\end{abstract}


\IEEEdisplaynontitleabstractindextext

%
\IEEEpeerreviewmaketitle

\section{Introduction}
%
%
%
%
\IEEEPARstart{O}{nline} Social Networks (OSNs) are essential platforms for people to interact with each other, communicate information, and spread social influence.
According to the Pew Research Center's report~\cite{auxier2021social}, about $70\%$ of Americans were on Facebook in 2021, and seven in ten of them visited the site daily.
However, OSNs are under threat from Sybil attacks, which create fake accounts (also called Sybils) on OSNs and use them for various malicious activities, such as distributing spam, phishing URLs, and malware, and manipulating public opinion and the stock market by spreading fake news.
For example, Sybils have been exploited to propagate anti-vaccine messages~\cite{broniatowski2018weaponized, allem2018could} and manipulate online political discussions~\cite{bessi2016social, bastos2019brexit}.
Therefore, Sybil detection is a fundamental task for OSN security.

The common Sybil detection approach is the graph-based approach that detects Sybils on the basis of the graph structure of OSNs (i.e., the friendship relation between users on OSNs).
This approach is motivated by the following observation: 
Sybils tend to be densely connected to other Sybils and sparsely connected to benign users because malicious attackers can easily control the connection between Sybils while they cannot control the connection between Sybils and benign users~\cite{alvisi2013sok}.
Therefore, it is expected that one can distinguish between a Sybil region and a benign region by exploiting the graph structure of OSNs.

Most graph-based methods predict unknown node labels (Sybil or benign) by assigning a prior reputation score to each node using known node labels and then updating and propagating the reputation score locally on a graph.
Two kinds of propagation algorithms are often used: random walk-based and loopy belief propagation-based.
Random walk-based methods~\cite{yang2012analyzing, cao2012aiding, boshmaf2016integro, jia2017random} propagate the trust or badness score by random walks from known benign or Sybil nodes and rank the Sybil-likeness of unknown nodes.
Loopy belief propagation methods~\cite{gong2014sybilbelief, fu2017robust, gao2018sybilfuse, dorri2018socialbothunter, wang2018structure} model the OSN structure as a pairwise Markov random field and compute the marginal distribution for each node (i.e., the probability that a node is Sybil) by a loopy belief propagation algorithm or its approximation.

However, although various Sybil detection methods have been proposed over the past decade, almost all methods have been compared just experimentally in terms of detection performance and noise robustness.
Since experimental results often depend on experimental conditions, such as dataset properties and experimental settings, a good result for an experimental condition does not guarantee the same for other ones.
To understand why and under what conditions each method works well, we need to compare them theoretically.
To this end, in our previous work~\cite{furutani2020sybil}, we formulated the random walk with restart and the loopy belief propagation algorithm as low-pass filtering and attempted a theoretical comparison of the performance of random walk-based and loopy belief propagation-based Sybil detection methods. 
However, this work does not provide a comprehensive comparison of existing detection methods (only a comparison between CIA~\cite{yang2012analyzing} and SybilBelief~\cite{gong2014sybilbelief}), nor can it explain the differences in detection performance for differences in structural properties of graphs (such as degree heterogeneity and modularity).

In this study, extending our previous work~\cite{furutani2020sybil}, we show that existing representative graph-based Sybil detection methods (CIA~\cite{yang2012analyzing}, SybilRank~\cite{cao2012aiding}, SybilWalk~\cite{jia2017random}, SybilBelief~\cite{gong2014sybilbelief}, and SybilSCAR~\cite{wang2018structure}) can be interpreted in a unified framework of low-pass filtering.
This framework enables us to theoretically compare and analyze each method from two perspectives: filter kernel properties and the spectrum of shift matrices.
Our analysis reveals that the detection performance of each method depends on how well low-pass filtering can extract low frequency components and remove noisy high frequency components.
In other words, for a Sybil detection method to perform well, 
1) the filter kernel must properly emphasize (remove) low (high) frequency components, and 
2) the low frequency eigenvectors of the shift matrix must have high community detectability.
Furthermore, on the basis of the analysis, we propose a novel detection method, called SybilHeat, with the filter kernel and the shift matrix that satisfies the above two requirements.
Our main contribution are summarized as follows:
\begin{itemize}
    \item We present the low-pass filtering framework for theoretically comparing and analyzing Sybil detection methods and identify the requirements for high performance of a Sybil detection method.
    \item We propose a Sybil detection method called SybilHeat that performs consistently better than other methods on graphs with various structural properties.
    \item We demonstrate the validity of our analysis and the performance of the proposed detection method through numerical experiments on synthetic graphs generated by the stochastic block model (SBM) and real social networks.
\end{itemize}

The rest of this paper is organized as follows:
In Section~II and Section~III, we present related work and preliminaries, respectively.
We interpret the existing methods as low-pass filtering in Section~IV.
In Section~V, we provide a theoretical comparison of the existing methods based on the interpretation and discuss our proposed method, SybilHeat.
In Section~VI, we evaluate the validity of our analysis and the performance of SybilHeat through numerical experiments.
Finally, Section~VII concludes our paper.

\section{Related Work}
In this section, we give a brief overview of Sybil detection methods.
Random walk-based methods~\cite{yu2006sybilguard, yu2008sybillimit, danezis2009sybilinfer, yang2012analyzing, cao2012aiding, boshmaf2016integro, jia2017random} detect Sybils by random walks from known labeled nodes on a graph.
SybilGuard~\cite{yu2006sybilguard} and SybilLimit~\cite{yu2008sybillimit} detect Sybils using special random walks called random routes.
In a normal random walk, the destination of a walker is randomly chosen for each step, whereas in random routes, the destination is predetermined by the permutation $\pi_v$ for each node $v$.
That is, random routes that enter from an edge $e$ always exit from edge $\pi_v(e)$.
An unlabeled node is approved as a benign node when the random routes originating from it intersect with the random routes from a known benign node.
SybilInfer~\cite{danezis2009sybilinfer} builds a probabilistic model of benign regions and uses it to detect potential Sybil regions.
SybilGuard, SybilLimit, and SybilInfer are not scalable to large OSNs and are not robust to label noise because they only use information from known benign nodes.
CIA~\cite{yang2012analyzing} propagates the badness score of each node by random walks with restart from known Sybil nodes.
SybilRank~\cite{cao2012aiding} evaluates the trust score of each node by computing the landing probability of early-terminated random walks from known benign nodes.
\'{I}ntegro~\cite{boshmaf2016integro} improves SybilRank by learning the edge weights and then considering random walks on the weighted graph.
The random walk-based method described above has the limitation that only labeled benign nodes or labeled Sybil nodes can be used (not both).
To overcome this problem, SybilWalk~\cite{jia2017random} computes the badness score of each node by random walks on the augmented graph with two additional nodes (Sybil label node and benign label node).

Loopy belief propagation methods~\cite{gong2014sybilbelief, fu2017robust, gao2018sybilfuse, dorri2018socialbothunter, wang2018structure} model the OSN structure as a pairwise Markov random field and compute the marginal distribution for each node (i.e., the probability that a node is Sybil) by a loopy belief propagation algorithm or its approximation.
SybilBelief~\cite{gong2014sybilbelief} first assigns the prior probability to each node using known node labels and then uses loopy belief propagation to calculate the posterior probability of them.
Later studies~\cite{fu2017robust, gao2018sybilfuse, dorri2018socialbothunter} have demonstrated that learning and exploiting node and edge features improve the performance of Sybilbelief.
SybilBelief and its variants rely on loopy belief propagation for inference, which is not scalable and has no convergence guarantees.
Wang \textit{et al.}~\cite{wang2018structure} provided a general framework that integrates random walk-based and loopy belief propagation-based methods and proposed SybilSCAR, a random walk-like score propagation algorithm, by approximating the loopy belief propagation algorithm.
SybilSCAR is more scalable than SybilBelief, and convergence is guaranteed.
However, this framework does not provide theoretical insight into the performance of existing Sybil detection methods.

Other Sybil detection methods include behavior-based detection methods~\cite{wang2010don, lee2010uncovering, yang2011free, amleshwaram2013cats, wang2015making, zheng2015detecting}.
They often use machine learning to classify users into benign or Sybil on the basis of their social behavior.
Most of them consist of two steps: 1) extracting behavior-based features that contribute to Sybil detection (e.g., tweet content and timing, follower/followee information, etc.), and then 2) constructing a detection model using the extracted features.
A major limitation of behavior-based methods is that attackers can easily imitate the behavior of benign users, thereby compromising the effectiveness of the method.

\section{Preliminaries}
\subsection{Graph Signal Processing}
In this subsection, we briefly introduce the basic concepts of graph signal processing~\cite{shuman2013emerging, ortega2018graph}.
Let $G=(V, E)$ be an unweighted undirected graph without self-loops and multiple edges, where $V=\{1, 2, \dots, N\}$ is the node set and $E \subset V \times V$ is the edge set.
A graph signal $x: V \to \mathbb{R}$ is the real-valued function defined on the node set $V$ and is represented as $N$-dimensional vector $\bm{x}=(x_1, x_2, \dots, x_N)$.
A shift matrix $\bm{S}=[S_{ij}] \in \mathbb{R}^{N \times N}$ is a matrix such that the off-diagonal element $S_{ij} \neq 0$ iff $(i,j) \not\in E$.
When the graph signal $\bm{x}$ is multiplied by the shift matrix $\bm{S}$, each element of the shifted signal $\tilde{\bm{x}}=\bm{S}\bm{x}$ is a linear combination of the signal value of its adjacent nodes, that is, the original graph signal is shifted over the graph.
In general, the adjacency matrix and Laplacian matrix are often used as the shift matrix~\cite{shuman2013emerging, ortega2018graph}.
The adjacency matrix $\bm{A}=[A_{ij}] \in \mathbb{R}^{N \times N}$ is a real symmetric matrix defined as $A_{ij}=1$ if $(i,j) \in E$ and $A_{ij}=0$ otherwise.
The Laplacian matrix is defined as $\bm{L}:=\bm{D}-\bm{A}$ where $\bm{D}:=\diag(d_{1},d_2,\dots,d_{N})$ is the degree matrix and $d_{i}:=\sum_{i=1}^{N}A_{ij}$ is node $i$'s degree.

We define the diagonal matrix $\bm{\Lambda} := \diag(\lambda_1, \lambda_2, \dots, \lambda_N)$ with eigenvalues $\lambda_1 \le \lambda_2 \le \cdots \le \lambda_N$ of $\bm{S}$ and the matrix $\bm{V} = (\bm{v}_1, \bm{v}_2, \dots, \bm{v}_N)$ with the eigenvector $\bm{v}_\mu$ corresponding to $\lambda_\mu$.
The graph Fourier transform (GFT) of $\bm{x}$ is defined as $\hat{\bm{x}} := \bm{V}^{-1}\bm{x}$ and inverse GFT is $\bm{x} := \bm{V}\hat{\bm{x}}$.
The graph filtering (also called graph convolution) of input signal $\bm{x}_{\mathrm{in}}$ is defined as
\begin{align}
    \bm{x}_{\mathrm{out}} = \bm{V} h(\bm{\Lambda}) \bm{V}^{-1} \bm{x}_{\mathrm{in}},
\end{align}
where $h(\bm{\Lambda}) := \diag(h(\lambda_1), h(\lambda_2), \dots, h(\lambda_N))$ and $h(\lambda)$ is a filter kernel function defined on the region $[\lambda_1, \lambda_N]$.
As with filtering in the classical signal processing, the graph filtering operation is interpreted as transforming the graph signal into the frequency domain signal by GFT, multiplying filter $h(\lambda)$, and then transforming back into the graph signal by the inverse GFT.
This outputs a signal in which specific frequency components of the input signal are amplified or attenuated.

\subsection{Stochastic Block Model}
A typical structural feature of real-world social networks is the existence of community structure.
Roughly speaking, the community is a group of nodes that are densely connected within a group and sparsely connected between groups.
One of the most basic models for generating random graphs with communities is the SBM~\cite{holland1983stochastic}.

Denoting $k$ communities by $\mathscr{C}_1, \mathscr{C}_2, \dots, \mathscr{C}_k$, SBM assumes node $i$ and node $j$ are connected with the probability
\begin{align}
    \mathrm{Pr}(A_{ij}=1) = \frac{C_{l(i), l(j)}}{N},
\end{align}
where the symmetric matrix $\bm{C}=[C_{ab}] \in \mathbb{R}^{k \times k}$ is the connectivity matrix and $C_{ab}/N$ is the probability of the edge being connected between nodes belonging to $\mathscr{C}_a$ and $\mathscr{C}_b$.
The map $l:V \to \{1, 2, \dots, k\}$ assigns a community to each node.
As a special case, let us consider the SBM with two symmetric communities ($|\mathscr{C}_1|=|\mathscr{C}_2|=N/2$) and denote $C_{ab}=c_{\mathrm{in}}$ if $a=b$ and $C_{ab}=c_{\mathrm{out}}$ if $a \neq b$.
In this case, it was conjectured in~\cite{decelle2011asymptotic} that two communities are detectable if and only if the inequality \begin{align}
    \frac{c_{\mathrm{in}} - c_{\mathrm{out}}}{2} > \sqrt{ \frac{c_{\mathrm{in}} + c_{\mathrm{out}}}{2} }
    \label{eq:SBM_detectable_condition}
\end{align}
holds, and it was proved in~\cite{mossel2012stochastic, massoulie2014community}.

The main limitation of the SBM is that all nodes within each community have the same average degree.
Degree-Corrected SBM (DCSBM)~\cite{karrer2011stochastic} is a more realistic model, which takes into account the degree heterogeneity of nodes within a community.
DCSBM connects node $i$ and node $j$ with the probability
\begin{align}
    \mathrm{Pr}(A_{ij}=1) = \theta_i\theta_j\frac{C_{l(i), l(j)}}{N},
\end{align}
where $\theta_i$ is the intrinsic connectivity of node $i$.
For each node $i$, $\theta_i$ is randomly sampled from the distribution $p(\theta)$ with $\mathbb{E}[\theta]=1$ and $\mathbb{E}[\theta^2]=\Phi$.
The intrinsic connectivity is proportional to the expected node degree (i.e., $\mathbb{E}[d_i] \propto \theta_i$) and produces an arbitrary degree distribution.
DCSBM includes SBM as a special case of $\forall i.~\theta_i=1$.
In~\cite{gulikers2018impossibility}, the detectable condition~(\ref{eq:SBM_detectable_condition}) for SBM is generalized to DCSBM as
\begin{align}
    \frac{c_{\mathrm{in}} - c_{\mathrm{out}}}{2} >
    \sqrt{ \frac{c_{\mathrm{in}} + c_{\mathrm{out}}}{2 \Phi} }.
    \label{eq:DCSBM_detectable_condition}
\end{align}

\section{Interpretation of Sybil Detection as Low-Pass Filtering}
\label{sect:4}
In this section, we explain how to interpret existing graph-based Sybil detection methods as low-pass filtering.
For an undirected graph $G=(V, E)$, let $V_s \subset V$ be the set of labeled Sybil nodes and $V_b \subset V$ be the set of labeled benign nodes.
Given a prior reputation score $\bm{q}=(q_1, q_2, \dots, q_N)^\TT$ and a graph $G$, existing graph-based Sybil detection methods can be understood as methods that iteratively update the reputation score $\bm{p}^{(t)}=(p_1^{(t)}, p_2^{(t)}, \dots, p_N^{(t)})^\TT$ at step $t$ following a certain update rule 
\begin{align}
    \bm{p}^{(t)} = f \! \left(\bm{p}^{(t-1)}; \bm{q}, G \right)
    \label{eq:updating_function}
\end{align}
until convergence, and then predict a label of each node using the final score $\bm{p}=\lim_{t \to \infty}\bm{p}^{(t)}$~\cite{wang2018graph}.
The prior reputation score $\bm{q}$ and the update rule $f(\cdot)$ differ from method to method.

We here consider reformulating~(\ref{eq:updating_function}) as the low-pass filtering
\begin{align}
    \bm{p} = \bm{V} h(\bm{\Lambda}) \bm{V}^{-1} \bm{q},
    \label{eq:low-pass}
\end{align}
where $\bm{\Lambda}$ and $\bm{V}$ are the diagonal matrix of eigenvalues of a shift matrix $\bm{S}$ and the invertible matrix consisting of its eigenvectors, respectively.
$h(\cdot)$ is the low-pass filter kernel.
This formulation gives a low-pass filtering interpretation to the existing Sybil detection methods and enables a theoretical comparison between them.
Hereafter, we describe the low-pass filtering interpretation of the following representative Sybil detection methods: CIA~\cite{yang2012analyzing}, SybilRank~\cite{cao2012aiding}, SybilWalk~\cite{jia2017random}, SybilBelief~\cite{gong2014sybilbelief}, and SybilSCAR~\cite{wang2018structure}.
Note that, for simplicity, we here consider unweighted undirected graphs, but our approach is easy to extend to weighted ones.

\subsection{CIA}
CIA~\cite{yang2012analyzing} propagates the badness score of each node by random walks with restart from labeled Sybil nodes $V_s$.
For a restart parameter $0 < \alpha < 1$, the update rule is given by
\begin{align}
    \bm{p}^{(t)} = \alpha \bm{A}\bm{D}^{-1}\bm{p}^{(t-1)} + (1 -\alpha) \bm{p}^{(0)}.
    \label{eq:CIA}
\end{align}
The initial score of node $i$ is set to $p_i^{(0)}=1$ if $i \in V_s$ and $p_i^{(0)}=0$ if $i \not\in V_s$.
Denoting $\bm{q}=\bm{p}^{(0)}$, we have the fixed point $\bm{p}$ of~(\ref{eq:CIA}) as
\begin{align}
    \bm{p} 
    &= (1-\alpha)(\bm{I} - \alpha \bm{A}\bm{D}^{-1})^{-1} \bm{q} \notag \\
    &= (1-\alpha)(\bm{I} - \alpha (\bm{I} - \Lrw) )^{-1} \bm{q} \notag\\
    &= \bm{V}_{\mathrm{r}} (1-\alpha)(\bm{I} - \alpha (\bm{I} - \bm{\Lambda}_{\mathrm{r}}) )^{-1} \bm{V}_{\mathrm{r}}^{-1} \bm{q},
    \label{eq:CIA_lowpass}
\end{align}
where $\bm{\Lambda}_{\mathrm{r}}$ and $\bm{V}_{\mathrm{r}}$ are matrices consisting of eigenvalues and eigenvectors of the random walk Laplacian $\Lrw := \bm{I} - \bm{A}\bm{D}^{-1}$, respectively.

\subsection{SybilRank}
SybilRank~\cite{cao2012aiding} evaluates the trust score of each node by computing the landing probability of early-terminated random walks from labeled benign nodes $V_b$.
This is motivated by the hypothesis that since the connection between Sybil and benign nodes is sparse, a random walk starting from a benign node and terminating in a finite step is less likely to reach a Sybil node, and thus the landing probability is higher for benign nodes and lower for Sybil nodes.
Setting the initial score to $p_i^{(0)}=1/|V_b|$ if $i \in V_b$ and $p_i^{(0)}=0$ if $i \not\in V_b$, the trust score $\bm{p}^{(t)}$ is updated by
\begin{align}
    \bm{p}^{(t)} = \bm{A}\bm{D}^{-1}\bm{p}^{(t-1)}.
    \label{eq:SybilRank}
\end{align}
SybilRank calculates the final trust score by terminating the above update equation at a finite step $\Gamma=O(\log N)$ and then normalizing the trust score by the degree to eliminate the degree bias (i.e., $p_i = p_i^{(\Gamma)}/d_i$).
Since $\bm{p}^{(\Gamma)} = (\bm{A}\bm{D}^{-1})^\Gamma \bm{p}^{(0)}$, we have the final trust score
\begin{align}
    \bm{p} 
    = \bm{D}^{-1}(\bm{I} - \Lrw)^\Gamma\bm{q}
    = \bm{D}^{-1} \bm{V}_{\mathrm{r}}^{} (\bm{I} - \bm{\Lambda}_{\mathrm{r}})^\Gamma \bm{V}_{\mathrm{r}}^{-1} \bm{q},
    \label{eq:SybilRank_lowpass}
\end{align}
with $\bm{q}=\bm{p}^{(0)}$.
Therefore, SybilRank can be interpreted as the operation combining the low-pass filtering by $\Lrw$ and degree-normalization.

\subsection{SybilWalk}
SybilWalk~\cite{jia2017random} computes the badness score of each node by random walks on the augmented graph $\widehat{G}=(V \cup \{l_s, l_b\}, \widehat{E})$ added two label nodes (Sybil label node $l_s$ and benign label node $l_b$) to an original graph $G$.
For the augmented graph $\widehat{G}$, label nodes $l_s$ and $l_b$ are respectively connected to known Sybil nodes and known benign nodes (i.e., $\widehat{E} = E \cup \{(i, l_b) \,|\, i \in V_b\} \cup \{(i, l_s) \,|\, i \in V_s\}$).
The badness score for each node $i \in V$ is calculated as the probability that a random walk starting from node $i$ will reach $l_s$ before reaching $l_b$ as follows:
\begin{align}
    p_i^{(t)} = \sum_{j = 1}^N \frac{a_{ij}}{\widehat{d}_i}p_j^{(t-1)},
    \label{eq:SybilWalk}
\end{align}
where $\widehat{d}_i$ is the degree of node $i$ in the augmented graph $\widehat{G}$ (i.e., $\widehat{d}_i=d_i + 1$ if $i \in V_b \cup V_s$ and $\widehat{d}_i = d_i$ otherwise).
The badness scores of label nodes are given by $p_{l_b}=0$ and $p_{l_s}=1$.

Indeed, SybilWalk is equivalent to an absorbing Markov chain with $l_b$ and $l_s$ as absorbing nodes.
For a random walk on $\widehat{G}$, the transition matrix between user nodes is $\widehat{\bm{D}}^{-1}\! \bm{A} \in \mathbb{R}^{N \times N}$ where $\widehat{\bm{D}}:=\diag(\widehat{d}_{1},\widehat{d}_2,\dots,\widehat{d}_{N})$, and the transition matrix from user nodes to label nodes is given by $\bm{Q} = (\bm{q}_b, \bm{q}_s) \in \mathbb{R}^{N \times 2}$.
Here, each component of $\bm{q}_s$ is defined as $q_{si}=1/\widehat{d}_i$ if $i \in V_s$ and $q_{si}=0$ if $i \not\in V_s$, and $\bm{q}_b$ is defined in the same way.
Hence,~(\ref{eq:SybilWalk}) is rewritten as $\bm{p}^{(t)} = \bm{\Pi}\bm{p}^{(t-1)}$ by using the transition matrix
\begin{align*}
    \bm{\Pi} := 
    \left(
        \begin{array}{c|c}
            \widehat{\bm{D}}^{-1}\!\bm{A} & \bm{Q} \\
            \hline
            \bm{O} & \bm{I}_2
        \end{array}
    \right),
\end{align*}
where $\bm{I}_2$ is the $2 \times 2$ identity matrix.
Therefore, we have
\begin{align}
    \bm{p} 
    &= \lim_{t \to \infty} 
    \bm{\Pi}^t \bm{p}^{(0)} = 
    \left(
        \begin{array}{c|c}
            \bm{O} & (\bm{I} - \widehat{\bm{D}}^{-1}\!\bm{A})^{-1}\bm{Q} \\
            \hline
            \bm{O} & \bm{I}_2
        \end{array}
    \right)
    \begin{pmatrix}
        \vdots \\
        0 \\
        1
    \end{pmatrix} \notag \\
    &= (\bm{I} - \widehat{\bm{D}}^{-1}\!\bm{A})^{-1}\bm{q}_s
    = \bm{V}_{\mathrm{a}}^{} \,\bm{\Lambda}_{\mathrm{a}}^{-1} \bm{V}_{\mathrm{a}}^{-1} \bm{q}_s,
    \label{eq:SybilWalk_lowpass}
\end{align}
where $\bm{\Lambda}_{\mathrm{a}}$ and $\bm{V}_{\mathrm{a}}$ are matrices consisting of eigenvalues and eigenvectors of the augmented normalized Laplacian $\Laug := \bm{I} - \widehat{\bm{D}}^{-1}\bm{A}$, respectively.

\subsection{SybilBelief}
SybilBelief~\cite{gong2014sybilbelief} models the OSN structure as a pairwise Markov random field and computes the marginal distribution for each node (i.e., the probability that a node is Sybil) by the standard loopy belief propagation~\cite{pearl1988probabilistic}.
Let us associate a random variable $s_i \in \{-1, +1\}$ to each node $i \in V$.
$s_i = +1$ means that node $i$ is Sybil, and $s_i=-1$ means that it is benign.
The pairwise Markov random field is defined as
\begin{align}
    p(s_1, s_2, \dots, s_N) = \frac{1}{Z} \prod_{(i,j) \in E}\psi_{ij}(s_i, s_j) \prod_{i \in V} \phi_i(s_i),
    \label{eq:pMRF}
\end{align}
where $Z$ is a normalization constant (called partition function), and $\phi_i(s_i)$ and $\psi_{ij}(s_i, s_j)$ are node and edge potential functions defined as follows, respectively:
\begin{align}
    \label{eq:node_potential_function}
    \phi_i(s_i) &= 
    \begin{cases}
        q_i & \text{if $s_i = +1$}\\
        1 - q_i & \text{if $s_i = -1$}
    \end{cases}
    ,\\
    \psi_{ij}(s_i, s_j) &= 
    \begin{cases}
        w_{ij} & \text{if $s_is_j = +1$}\\
        1 - w_{ij} & \text{if $s_is_j = -1$}
    \end{cases},
    \label{eq:edge_potential_function}
\end{align}
where $\vec{E}$ is the set of oriented edges of $E$ and satisfies $|\vec{E}|=2|E|$.
We can determine whether a node $i$ is Sybil or not by evaluating the marginal distribution $p_i(s_i)$.
However, it is exponentially hard to compute the marginal distribution directly from the joint distribution in~(\ref{eq:pMRF}).
The loopy belief propagation algorithm is a common method to calculate an approximate marginal distribution $b_i(s_i) \approx p_i(s_i)$.
This algorithm iteratively updates the probability distribution (called message) $\mu_{ij}(s_j)$ for each directed edge $(i,j) \in \vec{E}$.
The message from node $i$ to node $j$ at step $t+1$ is given by
\begin{align}
    \mu_{ij}^{(t+1)}(s_{j}) = \frac{1}{Z_{ij}} \sum_{s_{i}=\pm 1} \phi_{i}(s_{i}) \, \psi_{ij}(s_{i}, s_{j}) \prod_{k \in \partial_{i \setminus j}} \mu_{ki}^{(t)}(s_{i}),
  \label{eq:BP_message}
\end{align}
where $\partial_{i \setminus j} := \partial i \setminus \{j\}$ is the set of neighbors of node $i$ excluding recipient node $j$.
By using a converged message $\mu_{ij}^\infty$, the approximate marginal distribution $b_i(s_i)$ is computed as
\begin{align}
    b_{i}(s_i) = \frac{1}{Z_i} \phi_{i}(s_i) \prod_{k \in \partial i} \mu_{ki}^{\infty} (s_i).
    \label{eq:approximate marginal}
\end{align}
If $G$ is a tree ($G$ has no loops), $b_i(s_i)$ is exactly equal to $p_i(s_i)$.
If $G$ has loops, $b_i(s_i)$ is not equal to $p_i(s_i)$ but often provides a good approximation~\cite{murphy2013loopy}.

Since the loopy belief propagation algorithm is nonlinear, we have to linearize~(\ref{eq:BP_message}) to represent it as~(\ref{eq:low-pass}).
In our previous study~\cite{furutani2020sybil}, we linearize~(\ref{eq:BP_message}) around a fixed point and reformulate SybilBelief as low-pass filtering by using a Bethe-Hessian matrix.
The Bethe-Hessian matrix is defined as
\begin{align}
    \bm{H}(r) := (r^2-1)\bm{I} + \bm{D} -r\bm{A},
    \label{eq:bethe-hessian}
\end{align}
with the parameter $r \in \mathbb{R}$.
When $r=1$, the Bethe-Hessian matrix becomes the Laplacian matrix $\bm{L}$.
Hereafter, unless otherwise noted, we set the parameter $r$ of the Bethe-Hessian $\bm{H}(r)$ is set to $r=[(\sum_{i}d_i^2)/(\sum_{i} d_i) - 1]^{1/2}$ as in~\cite{saade2014spectral}.
This setting has the advantage that informative eigenvalues are negative while the bulk of uninformative eigenvalues is positive, making them easy to distinguish between them.

The low-pass filtering interpretation of SybilBelief is given by
\begin{align}
    \bm{p} = \bm{V}_{\mathrm{H}}^{} \,g(\bm{\Lambda}_{\mathrm{H}}) \bm{V}_{\mathrm{H}}^{-1} \bm{q},
    \label{eq:SybilBelief_lowpass}
\end{align}
where $\bm{\Lambda}_{\mathrm{H}}$ and $\bm{V}_{\mathrm{H}}$ are matrices consisting of eigenvalues and eigenvectors of $\bm{H}(r)$, respectively.
The function $g(\lambda)$ is the ideal low-pass filter kernel; i.e., $g(\lambda)=1$ if $\lambda \le \lambda'$ and $g(\lambda) = 0$ otherwise.
For details on the derivation of~(\ref{eq:SybilBelief_lowpass}), see Appendix.

\subsection{SybilSCAR}
SybilBelief has the limitation of low scalability and no convergence guarantee of~(\ref{eq:BP_message}).
To overcome these limitations, SybilSCAR~\cite{wang2018structure} computes the probability $p_i$ of each node $i$ being Sybil by approximating~(\ref{eq:BP_message}) by replacing $\partial_{i \setminus j}$ with $\partial i$.
SybilSCAR assigns the prior probability $q_i$ to each node as
\begin{align*}
    q_i = 
    \begin{cases}
        0.5 + \theta & \text{if $i \in V_s$} \\
        0.5 - \theta & \text{if $i \in V_s$} \\
        0.5  & \text{otherwise} 
    \end{cases},
\end{align*}
where $\theta \in (0, 0.5]$ indicates assigning high prior probabilities to labeled Sybil nodes.
For a variable $y$, let us define a residual variable $\check{y} := y - 1/2$.
The update function of SybilSCAR is given by
\begin{align}
    \check{\bm{p}}^{(t)} = 2\check{\bm{W}}\check{\bm{p}}^{(t-1)} + \check{\bm{q}},
    \label{eq:SybilSCAR}
\end{align}
where $\check{\bm{W}}=(\check{w}_{ij})$ is a residual weight matrix with the element $\check{w}_{ij}=0$ if $(i,j) \not\in E$, and we set to $\check{\bm{p}}^{(0)} = \check{\bm{q}}$.

In~\cite{wang2018structure}, two SybilSCAR algorithms are proposed: SybilSCAR-C and SybilSCAR-D.
For an edge $(i,j) \in E$, SybilSCAR-C has a constant residual weight (i.e., $\check{w}_{ij} = 1/(2d_{\max})$), while SybilSCAR-D has a degree-normalized residual weight (i.e., $\check{w}_{ij} = 1/(2d_j)$).
For a fixed point $\check{\bm{p}}$ of~(\ref{eq:SybilSCAR}), we have $\check{\bm{p}} = 2\check{\bm{W}}\check{\bm{p}} + \check{\bm{q}} = (\bm{I} - 2\check{\bm{W}})^{-1}\check{\bm{q}}$, and thus SybilSCAR-C is rewritten as
\begin{align}
    \check{\bm{p}} 
    = \left(\bm{I} - \frac{1} { d_{\max}} \bm{A}\right)^{-1}\!\check{\bm{q}}
    = \bm{V}_{\mathrm{m}}^{} \bm{\Lambda}_{\mathrm{m}}^{-1} \bm{V}_{\mathrm{m}}^{-1} \check{\bm{q}},
    \label{eq:SybilSCAR-C_lowpass}
\end{align}
where $\bm{\Lambda}_{\mathrm{m}}$ and $\bm{V}_{\mathrm{m}}$ are matrices consisting of eigenvalues and eigenvectors of the maximum degree-normalized Laplacian $\Lmax := \bm{I} - \frac{1} { d_{\max}} \bm{A}$, respectively.
Also, SybilSCAR-D is rewritten as
\begin{align}
    \check{\bm{p}} 
    = \left(\bm{I} - \bm{A}\bm{D}^{-1}\right)^{-1}\!\check{\bm{q}}
    = \bm{V}_{\mathrm{r}}^{}\, \bm{\Lambda}_{\mathrm{r}}^{-1} \bm{V}_{\mathrm{r}}^{-1} \check{\bm{q}}.
    \label{eq:SybilSCAR-D_lowpass}
\end{align}

\begin{figure*}[!t]
    \centering
    \includegraphics[width=1.0\linewidth]{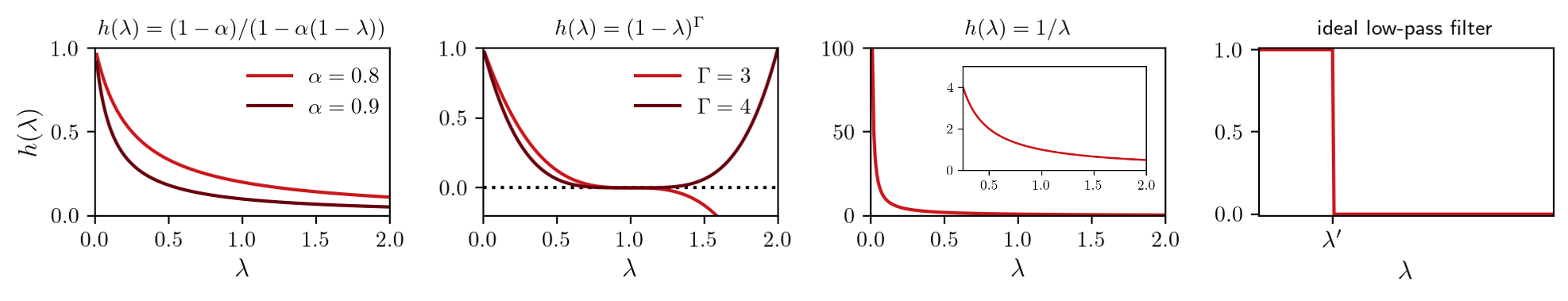}
    \caption{Low-pass filter kernels of existing Sybil detection methods}
    \label{fig:filter}
\end{figure*}

\begin{table}[tb]
 \centering
 \renewcommand{\arraystretch}{1.3}
 \caption{Summarization of low-pass filtering interpretation of representative Sybil detection methods}
  \begin{tabular}{llll}
   \hline
   Method & Shift matrix $\bm{S}$ & Filter kernel $h(\lambda)$ \\
   \hline
    CIA & $\Lrw $
    & $(1-\alpha)/(1-\alpha(1-\lambda))$ \\
    SybilRank & $\Lrw $
    & $(1-\lambda)^\Gamma$ \\
    SybilWalk
    & $\Laug $
    & $1/\lambda$ \\
    SybilSCAR-C & $\Lmax $
    & $1/\lambda$ \\
    SybilSCAR-D & $\Lrw $
    & $1/\lambda$ \\
    SybilBelief & $\bm{H}(r) $
    & ideal low-pass filter \\
   \hline
  \end{tabular}
  \label{tab:summary}
\end{table}

\section{Theoretical Comparisons}
\label{sect:5}
In Section~\ref{sect:4}, we described the low-pass filtering interpretation of some representative Sybil detection methods.
As shown in Table~\ref{tab:summary}, the differences between these methods can be attributed to the differences in their corresponding shift matrix and filter kernel.
The output of low-pass filtering depends on the property of the low-pass filter kernel and the choice of shift matrix (i.e., what Fourier basis is used for the frequency transform).
As is well known in the context of spectral clustering and graph signal processing, the eigenvectors corresponding to the small (low frequency) eigenvalues of the Laplacian contain rich information about the global community structure of a graph, while the eigenvectors corresponding to large (high frequency) eigenvalues contain noisy information~\cite{von2007tutorial, shuman2013emerging}.
Thus, the performance of the Sybil detection method is expected to depend on how well the low pass filtering can extract the low frequency components and remove the high frequency components of the input signal.
In this section, we compare and analyze each method from two perspectives: filter kernel properties and the spectrum of the shift matrix.
Furthermore, on the basis of the theoretical insights, we propose a novel Sybil detection method called SybilHeat.

\subsection{Filter Kernel Properties}
\label{sect:5.1}
Figure~\ref{fig:filter} plots the four different filter kernels in Table~\ref{tab:summary}.
First, the CIA filter kernel $h(\lambda) = (1-\alpha)/(1-\alpha(1-\lambda))$ does not remove high frequency components sufficiently, so the output signal may be affected by noisy high frequency components.
For this reason, CIA is expected to have poor detection performance and noise robustness.

Next, the SybilRank filter kernel $h(\lambda) = (1-\lambda)^\Gamma$ removes frequencies in the middle range ($0.5 \le \lambda \le 1.5$) but passes high frequency components ($\lambda > 1.5$).
Therefore, if the largest eigenvalue $\lambda_N$ takes a large value, SybilRank may be strongly affected by high frequency components.
As described below, since the largest eigenvalue $\lambda_N^{\mathrm{r}}$ of the random walk Laplacian tends to become larger for a sparse graph, SybilRank is expected to perform poorly on sparse graphs.

The filter kernel $h(\lambda) = 1/\lambda$ strongly emphasizes low frequency components, and thus the contribution of high frequency components is relatively small.
Since $h(\lambda) \to \infty$ for $\lambda \to 0$, the contribution of the eigenvector $\bm{v}_1$ corresponding to the smallest eigenvalue, which is uninformative in general, is dominant.
Particularly, since the smallest eigenvalue of $\Lrw$ is $0$, SybilSCAR-D may fail detection.
However, if low frequency eigenvalues and high frequency eigenvalues are sufficiently separated, that is, the eigengap between informative and uninformative eigenvalues, $|\lambda_k - \lambda_{k+1}|$, is sufficiently large, high detection performance and noise robustness are expected. 

The filter kernel corresponding to SybilBelief equally extracts the low frequency components and completely removes the high frequency components, by definition.
Therefore, SybilBelief is expected to exhibit high detection performance and noise robustness as long as the low frequency eigenvectors are informative.

\subsection{Spectrum of Shift Matrices}
The output of low-pass filtering for a graph signal depends on how the frequencies (eigenvalues) are distributed and what Fourier basis is used for frequency transformation (i.e., the choice of shift matrix).
We here discuss each method focusing on the spectrum (eigenvalues and eigenvectors) of the shift matrix.
\subsubsection{Eigenvalues of shift matrices}
\begin{figure*}[!t]
    \centering
    \includegraphics[width=1.0\linewidth]{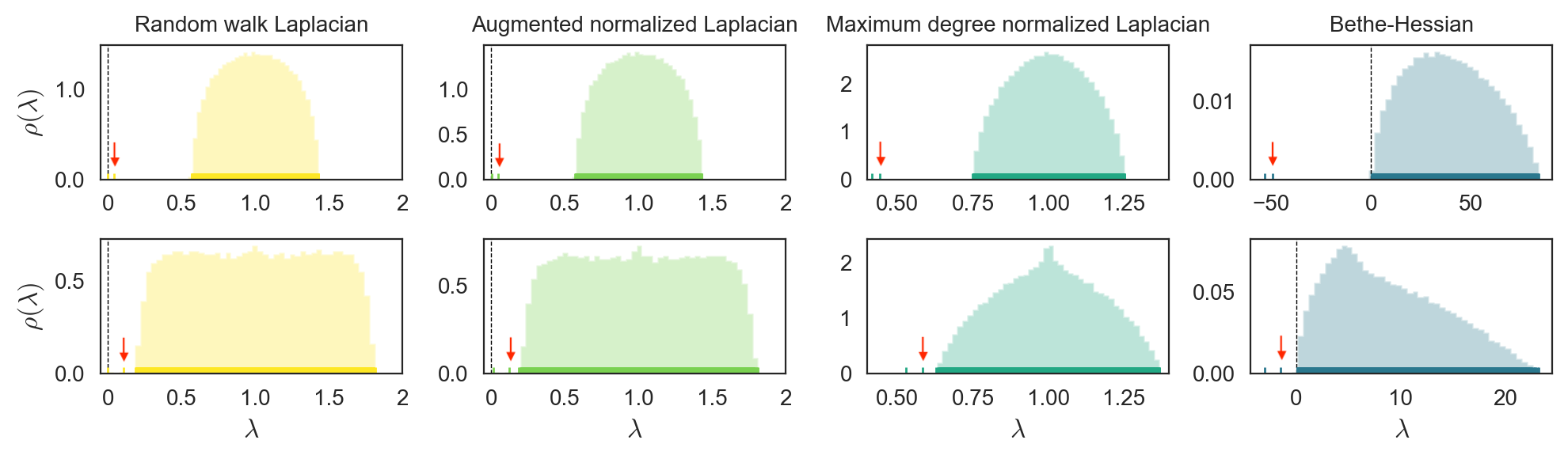}
    \caption{Spectral distributions of shift matrices for a dense modular graph generated by SBM with $N=3000$, $k=2$, $d_{\mathrm{ave}}=20$, $c_{\mathrm{out}}=1$ (upper panels) and a sparse modular graph generated by SBM with $N=3000$, $k=2$, $d_{\mathrm{ave}}=5$, $c_{\mathrm{out}}=1$ (lower panels).
    In each panel, the mark (``$\,|\,$'') stemming from the baseline is the position of eigenvalues, and the red arrow points to the second smallest eigenvalue $\lambda_2$.}
    \label{fig:spec_dist}
\end{figure*}

First, we discuss detection methods in terms of eigenvalues of the shift matrix.
Although frequencies are sampled evenly in classical signal processing, in graph signal processing, however, the frequency (eigenvalue) distribution is uneven and differs depending on the shift matrix.
Since the quality of the low-pass filtering is determined by how well it can extract low frequency components, it is anticipated that the more clearly low frequency eigenvalues are isolated from the bulk of high frequency eigenvalues on the eigenvalue distribution, the better the low-pass filtering is.
Figure~\ref{fig:spec_dist} shows the spectral distribution of shift matrices for a dense and sparse modular graph generated by SBM.
For dense modular graphs, $k$ small (low frequency) eigenvalues are clearly isolated from the bulk of high frequency eigenvalues for each shift matrix.
On the other hand, for sparse modular graphs, the bulk of uninformative eigenvalues is spread out, making them difficult to distinguish informative and uninformative eigenvalues.
This suggests that the high frequency components are difficult to sufficiently remove by low-pass filtering for sparse graphs, and thus the detection performance of any detection method will be worse than that of the dense graph case.

\subsubsection{Eigenvectors of shift matrices}
\label{sect:5.2.2}
Next, we discuss detection methods in terms of the eigenvectors of the shift matrix.
Since Sybil detection is essentially a problem of identifying Sybil and benign regions of a graph, the more informative the low frequency eigenvectors of each shift matrix are about the community structure of the graph, the better the detection performance is likely to be.
Therefore, to measure the informativeness of low frequency eigenvectors, we evaluate the community detectability of each shift matrix.
Specifically, we estimate communities by spectral clustering algorithm (Algorithm~\ref{alg:SC}) with each shift matrix of graphs with $k=2$ communities generated by SBM and DCSBM and compare the Normalized Mutual Information (NMI) score of true and estimated communities.

\begin{figure}[!t]
\begin{algorithm}[H]
  \caption{Spectral clustering algorithm}
  \label{alg:SC}
  \begin{algorithmic}[1]
    \REQUIRE Shift matrix $\bm{S}$, the number $k$ of communities
    \STATE Compute $k$ smallest eigenvectors of $\bm{S}$ and construct the $N \times k$ eigenvector matrix $\bm{V}_k = (\bm{v}_1, \bm{v}_2, \dots, \bm{v}_k)$
    \STATE Normalize the rows of $\bm{V}_k$
    \STATE For $i = 1, 2, \dots, N$, let $\bm{y}_i$ be the vector corresponding to the $i$-th row of $\bm{V}_k$
    \STATE Cluster the points $\{\bm{y}_i\}_{i=1}^N$ with $k$-means into $k$ communities
    \ENSURE Estimated communities $\hat{\mathscr{C}}_1, \hat{\mathscr{C}}_2, \dots, \hat{\mathscr{C}}_k$
  \end{algorithmic}
\end{algorithm}
\end{figure}

\begin{figure}[t]
    \centering
    \includegraphics[width=1.0\linewidth]{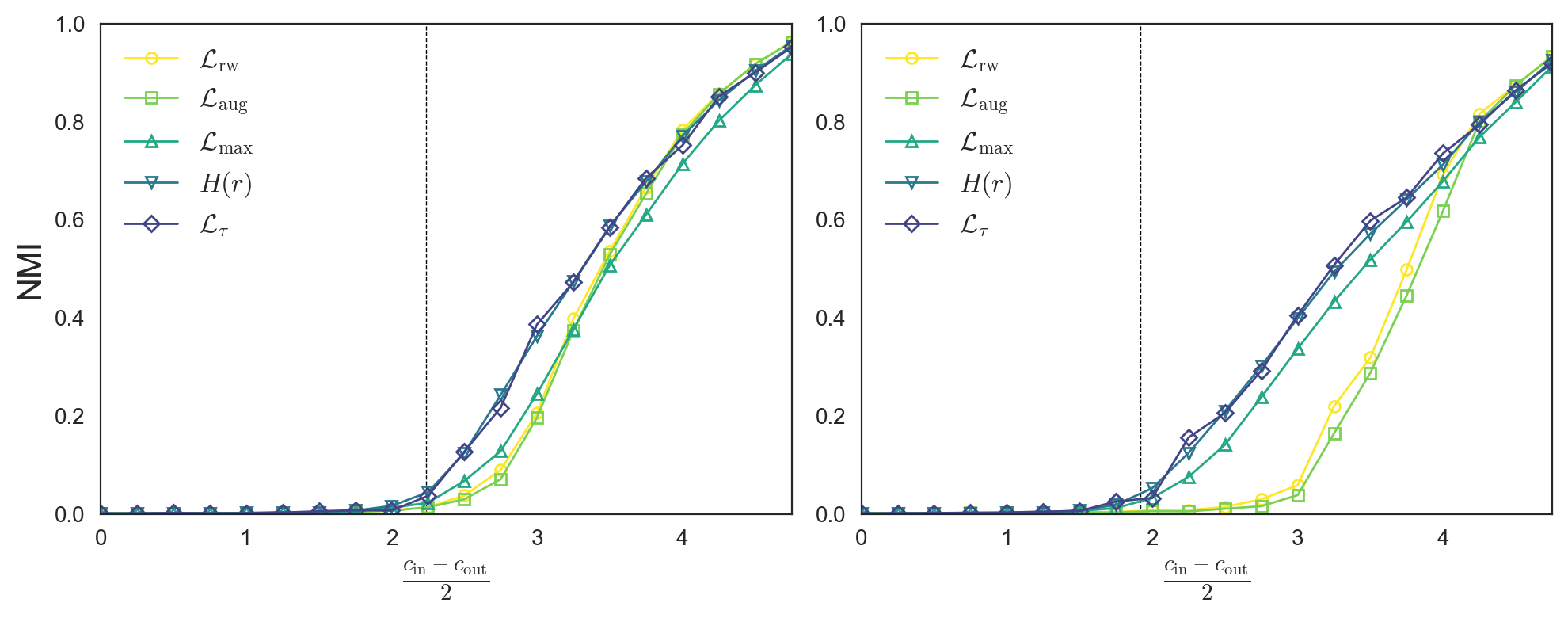}
    \caption{
    Community detectability of low frequency eigenvectors of the random walk Laplacian, augmented normalized Laplacian, maximum degree-normalized Laplacian, Bethe-Hessian, and regularized Laplacian for sparse modular graphs generated by SBM (left) and DCSBM (right).
    The vertical dotted line shows the detectability threshold.
    Parameters are $N = 1000$, $k=2$, $d_{\mathrm{ave}} = 5$, and $\theta_i=1$ for SBM and $\theta_i \sim [\mathcal{U}(3,7)]^3$ for DCSBM.
    Simulations are averaged over $100$ runs.}
    \label{fig:detectability}
\end{figure}

Figure~\ref{fig:detectability} shows the community detectability of low frequency eigenvectors of each shift matrix for sparse modular graphs generated by SBM and DCSBM.
The vertical dashed lines represent the detectability threshold on the right-hand side of equations~(\ref{eq:SBM_detectable_condition}) and~(\ref{eq:DCSBM_detectable_condition}), respectively.
First, for SBM graphs, all shift matrices can detect communities in the detectable region, and in particular, the Bethe-Hessian shows the highest detection performance.
For DCSBM graphs, the maximum degree-normalized Laplacian $\Lmax$ and the Bethe-Hessian $\bm{H}(r)$ can detect communities in the detectable region, while the random walk Laplacian $\bm{\mathcal{L}}_{\textrm{rw}}$ and the augmented normalized Laplacian $\bm{\mathcal{L}}_{\textrm{aug}}$ cannot detect the community when $(c_{\mathrm{in}}-c_{\mathrm{out}})/2$ is small (i.e., weakly modular) even in the detectable region.
The same is true for $k>2$ communities.
This suggests that SybilBelief and SybilSCAR-C perform well on sparse, degree heterogeneous, strong modular graphs.

To explain the reason for the high community detectability of $\Lmax$ and $\bm{H}(r)$, we have to review the regularization of a matrix.
In the previous studies~\cite{chaudhuri2012spectral, qin2013regularized}, a method called regularized spectral clustering (RSC) was proposed to improve spectral clustering by using the regularized Laplacian $\LL_\tau = \bm{I} - \bm{D}_{\tau}^{-1/2}\bm{A}\bm{D}_{\tau}^{-1/2}$ or $\LL_\tau = \bm{I} - \bm{D}_{\tau}^{-1}\bm{A}$ instead of the random walk Laplacian $\Lrw$.
Here, $\bm{D}_\tau := \bm{D} + \tau \bm{I}$ is a regularized degree matrix and $\tau \in \mathbb{R}_{\ge 0}$ is a regularization parameter.
For a graph generated by DCSBM, Qin \textit{et al.}~\cite{qin2013regularized} proved that the upper bound of the clustering error of RSC is proportional to $1/(d_{\min} + \tau)$ and $1/\bar{\lambda}_k^2$, where $\bar{\lambda}_k$ is the $k$ smallest eigenvalue of the expectation of $\LL_\tau$, $\mathscr{L}_\tau = \mathbb{E}[\LL_\tau]$.
For small $\tau$ (i.e., insufficient regularization), $1/(d_{\min} + \tau)$ becomes large, and conversely, for large $\tau$ (i.e., excessive regularization), $1/\bar{\lambda}_k$ becomes large.
Therefore, this result suggests that the clustering error of RSC can be minimized by setting appropriate $\tau$.
Qin \textit{et al.}~\cite{qin2013regularized} proposed $\tau = d_{\mathrm{ave}}$ as a suitable choice.
Indeed, as shown in Fig.~\ref{fig:detectability}, the community detectability of $\LL_\tau$ with $\tau = d_{\mathrm{ave}}$ is comparable to that of the Bethe-Hessian, which has the best performance.

Given the above, the results in Fig.~\ref{fig:detectability} can be explained as follows.
Defining the diagonal matrix $\widehat{\bm{I}}$ as $[\widehat{I}]_{ii}=1$ if $i \in V_b \cup V_s$ and $[\widehat{I}]_{ii}=0$ otherwise, the augmented normalized Laplacian is represented as $\Laug = \bm{I} - (\bm{D} + \widehat{\bm{I}})^{-1}\bm{A}$ and can be regarded as being insufficiently and unevenly regularized.
The maximum degree-normalized Laplacian can be regarded as being excessively and unevenly regularized since it can be rewritten as $\Lmax = \bm{I} - (\bm{D} + \bm{D}_{\mathrm{diff}})^{-1}\bm{A}$ with $[D_{\mathrm{diff}}]_{ii} := d_{\max} - d_i$.
From~(\ref{eq:bethe-hessian}), Bethe-Hessian has the following relationship with the regularized Laplacian:
\begin{align*}
    &\bm{H}(r) \bm{v} = \lambda \bm{v} \\
    \Leftrightarrow~ &(\bm{I} - r(\bm{D}+ (r^2 - \lambda - 1)\bm{I})^{-1}\bm{A})\bm{v} = 0 \\
    \Leftrightarrow~ &\bm{\mathcal{L}}_{r^2-\lambda-1} \bm{v} = \frac{r-1}{r}\bm{v}.
\end{align*}
Hence, the spectrum of the Bethe-Hessian and the regularized Laplacian are closely related.

\subsection{SybilHeat}
The above analysis reveals that the detection performance of each method depends on how well low-pass filtering can extract low frequency components and remove noisy high frequency components.
More specifically, for a Sybil detection method to perform well, 
1) the low-pass filter kernel $h(\lambda)$ must properly emphasize (remove) low
(high) frequency components, and 
2) low frequency eigenvectors of the shift matrix $\bm{S}$ must have high community detectability.
On the basis of this result, we propose a novel Sybil detection method (SybilHeat) with the filter kernel $h(\lambda) = e^{-s \lambda} ~(s \ge 0)$ and the shift matrix $\bm{S} = \LL_\tau = \bm{I} - \bm{D}_{\tau}^{-1/2}\bm{A}\bm{D}_{\tau}^{-1/2}~(\tau = d_{\mathrm{ave}})$ that satisfy the above two requirements.
The filter kernel $h(\lambda) = e^{-s \lambda}$ is called the heat kernel, and the larger the scaling parameter $s \ge 0$ is, the more strongly high frequency components are reduced.
The eigenvectors of $\LL_\tau$ have high community detectability comparable to the Bethe-Hessian, as described above.
For given prior reputation score $\bm{q}$, SybilHeat calculates the posterior reputation score $\bm{p}$ as
\begin{align}
    \bm{p} = \bm{V}_{\tau}e^{-s\bm{\Lambda}_{\tau}}\bm{V}_{\tau}^{-1}
    = e^{- s\LL_\tau} \bm{q}
    = h(\LL_\tau) \,\bm{q},
    \label{eq:sybilheat}
\end{align}
where $\bm{\Lambda}_{\tau}$ and $\bm{V}_{\tau}$ are matrices consisting of eigenvalues and eigenvectors of $\LL_\tau$, respectively.

Although all eigenvalues and eigenvectors are needed to naively compute $h(\LL_\tau)$, the time complexity of the eigenvalue decomposition is $O(N^3)$, which is not scalable for large $N$.
However, fortunately, the Chebyshev polynomial approximation~\cite{hammond2011wavelets} enables us to avoid computing the eigenvalue decomposition and approximately compute~(\ref{eq:sybilheat}).
Specifically, by using the (shifted) Chebyshev polynomials $\{\tilde{T}_k(\lambda)\}_{k=0}^\infty$ defined on the range $\lambda \in [0,2]$, we approximate $h(\LL_\tau) \bm{q}$ as
\begin{align}
    h(\LL_\tau) \bm{q}
    \approx \frac{\tilde{c}_0}{2} \bm{q} + \sum_{k=1}^K \tilde{c}_k \tilde{T}_k(\LL_\tau) \bm{q},
    \label{eq:cheb_approx}
\end{align}
where $K$ is the approximation order.
The Chebyshev coefficient $\tilde{c}_k~(k=0,1,\dots)$ is calculated by
\begin{align}
    \tilde{c}_k = \frac{2}{\pi}\int_0^\pi h(\cos\theta + 1) \cos(k\theta) d\theta.
\end{align}
By definition of the Chebyshev polynomials, $\tilde{T}_{0}(\LL_\tau)=\bm{I}$, $\tilde{T}_{1}(\LL_\tau)=\LL_\tau - \bm{I}$, and, for $k \ge 2$, $\tilde{T}_{k}(\LL_\tau)$ satisfies 
\begin{align*}
    \tilde{T}_k(\LL_\tau) = 2(\LL_\tau - \bm{I}) \tilde{T}_{k-1}(\LL_\tau)  - \tilde{T}_{k-2}(\LL_\tau).
\end{align*}
This indicates that a vector $\tilde{T}_{k}(\LL_\tau)\bm{q}$ in~(\ref{eq:cheb_approx}) can be recursively computed from $\tilde{T}_{k-1}(\LL_\tau)\bm{q}$ and $\tilde{T}_{k-2}(\LL_\tau)\bm{q}$.
Dominant in this computational cost is the matrix-vector multiplication of $\LL_\tau$, and its time complexity is $O(|E|)$~\cite{hammond2011wavelets}.
Thus, the overall time complexity of~(\ref{eq:cheb_approx}) is $O(K|E|)$, and SybilHeat is applicable to large social networks.

\section{Empirical Evaluations}
In the previous section, on the basis of the low-pass filtering interpretation of existing Sybil detection methods, we explained the reasons for the superiority or inferiority of the performance and the requirements for high performance of each method.
In addition, we proposed SybilHeat on the basis of our analysis.
In this section, we validate our analysis and evaluate the detection performance and noise robustness of SybilHeat through experiments on synthetic graphs and real-world social networks.

\subsection{Experimental Setup}
We use synthetic graphs and some real-world social networks for the experiments.
Synthetic graphs are generated by the SBM and DCSBM with average degree $d_{\mathrm{ave}}=5$ and $k=2$ even-sized communities.
We assume that one community is the benign region and the other is the Sybil region.
We also assume that $10\%$ of nodes randomly selected from the benign region are labeled benign nodes and $10\%$ of nodes randomly selected from the Sybil region are labeled Sybil nodes.

We also use real social networks with community structure that are benchmark dataset for community detection tasks for evaluating the detection performance and noise robustness: 
Zachary karate club~\cite{zachary1977information} (34 nodes and 78 edges), dolphin social network~\cite{lusseau2003bottlenose} (62 nodes and 159 edges), American colledge football~\cite{girvan2002community} (115 nodes and 613 edges), and political blogs~\cite{adamic2005political} (1224 nodes and 33430 edges).
For evaluation, we used the largest connected component of each graph, with half of the communities as benign regions and the rest as Sybil regions.
The $\max(3, \lfloor 0.1N \rfloor)$ nodes randomly selected from the benign region were labeled benign nodes, and the $\max(3, \lfloor 0.1N \rfloor)$ nodes randomly selected from the Sybil region were labeled Sybil nodes.

We compare SybilHeat with CIA, SybilRank, SybilWalk, SybilSCAR-C, and SybilBelief.
SybilSCAR-D is excluded because its convergence is not stable.
The experimental parameters for each method were set as: 
the restart parameter $\alpha=0.85$ for CIA, the number of iteration $\Gamma = \lfloor \log N \rfloor$ for SybilRank, the residual prior probability $\theta = 0.5$ for SybilSCAR, and the scaling parameter $s=8$ for SybilHeat.

\subsection{Detection Performance}
Since the Sybil detection method provides a ranking for each node such that Sybil nodes are ranked higher than benign nodes~\cite{viswanath2010analysis}, we adopt the Area Under the Receiver Operating Characteristic Curve (AUC) to evaluate detection performance.
The AUC of a method is the probability that a (randomly selected) Sybil node ranks higher than a benign node, and the AUC is 1.0 if all Sybil nodes rank higher than benign nodes.
If all nodes are ranked uniformly at random, the AUC is 0.5.

\begin{figure}[t]
    \centering
    \includegraphics[width=1.0\linewidth]{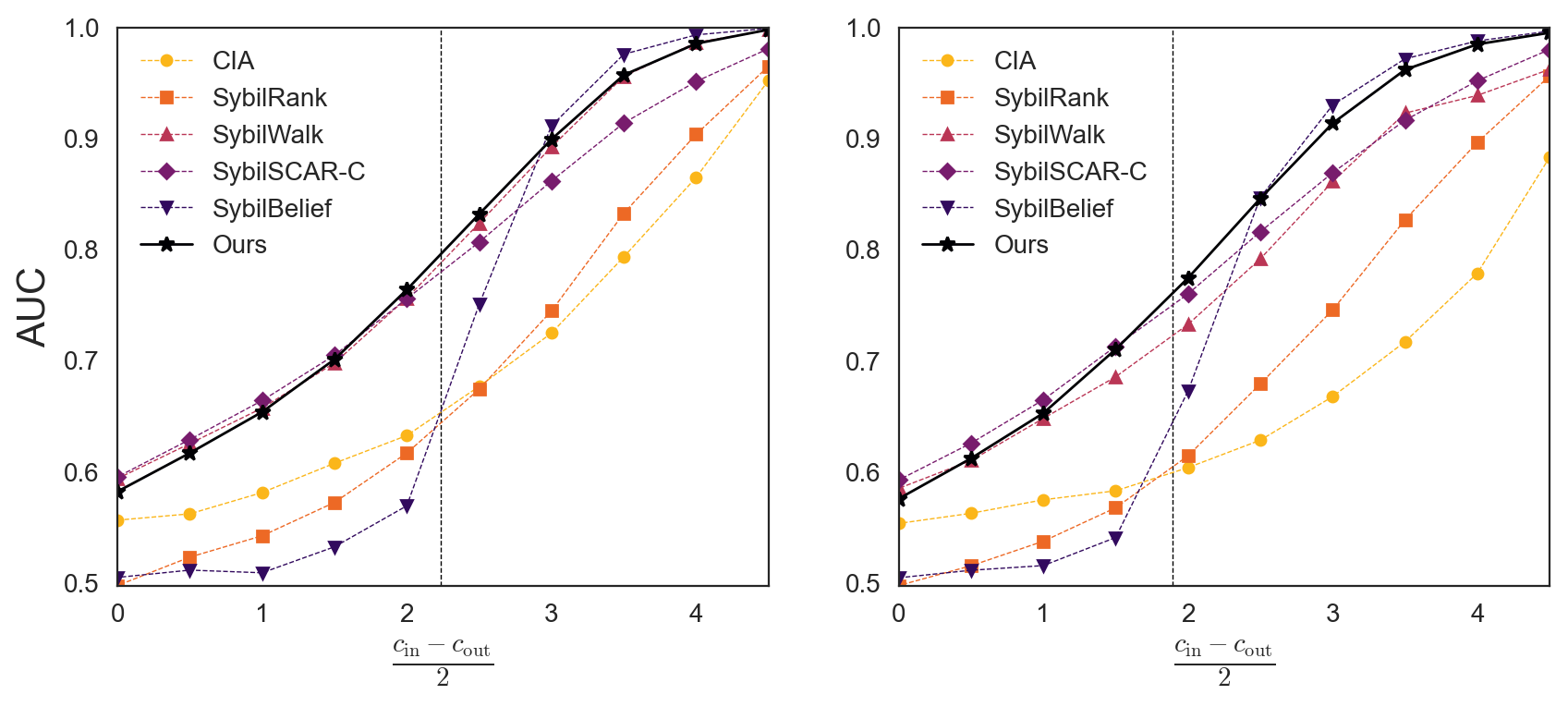}
    \caption{
    Detection performance of CIA, SybilRank, SybilWalk, SybilSCAR, SybilBelief, and SybilHeat for sparse modular graphs generated by SBM (left) and DCSBM (right).
    The vertical dotted line shows the detectability threshold.
    Parameters are $N = 1000$, $k=2$, $d_{\mathrm{ave}} = 5$, and $\theta_i=1$ for SBM and $\theta_i \sim [\mathcal{U}(3,7)]^3$ for DCSBM.
    Simulations are averaged over $100$ runs.}
    \label{fig:performance}
\end{figure}

Figure~\ref{fig:performance} shows the detection performance of each method with respect to the community strength $|c_{\mathrm{in}} - c_{\mathrm{out}}|/2$ on synthetic graphs generated by the SBM (left) and DCSBM (right).
We observe the following results from this figure.
First, the detection performance of all methods increases as the modularity (i.e., the difference between connectivity of intra- and inter-community) increases.
Second, random walk-based methods (CIA, SybilRank, and SybilWalk) perform worse for DCSBM graphs than for SBM graphs.
This is due to the low community detection performance of the shift matrices corresponding to these methods for DCSBM graphs, as shown in Fig.~\ref{fig:detectability}.
Moreover, SybilBelief shows the best performance in the detectable region, while in the undetectable region, its detection performance drops sharply. 
In other words, SybilBelief performs well only on strongly modular graphs. 
This is due to the fact that SybilBelief relies on only the $k$ small eigenvectors of the Bethe-Hessian (which have no information about the community structure in the undetectable region) to perform detection.
On the other hand, SybilHeat performs consistently better over the two regions than the other methods.
That is, SybilHeat performs better next to SybilBelief in the detectable region and performs best in the undetectable region, comparable to SybilWalk and SybilSCAR-C.

\begin{figure}[t]
    \centering
    \includegraphics[width=1.0\linewidth]{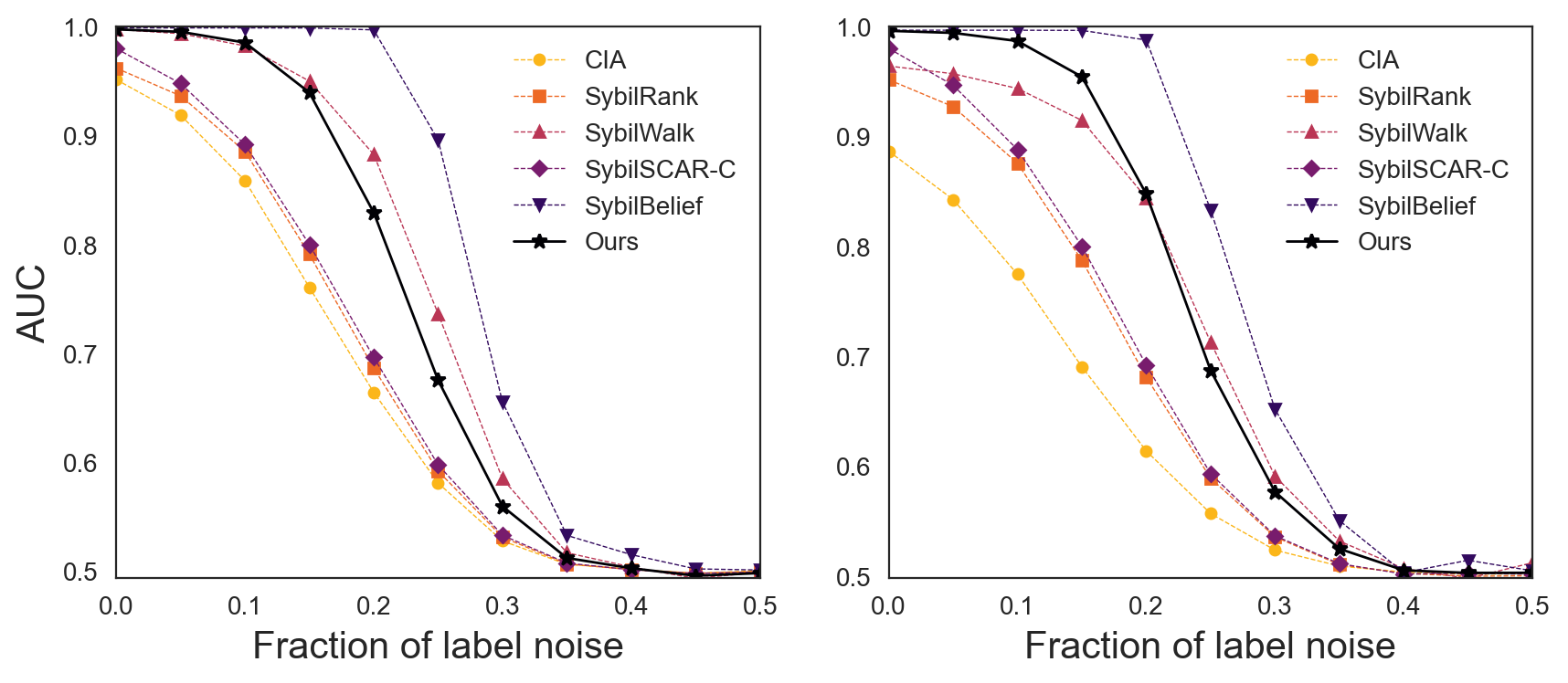}
    \caption{
    Noise robustness of CIA, SybilRank, SybilWalk, SybilSCAR, SybilBelief, and SybilHeat for sparse modular graphs generated by SBM (left) and DCSBM (right).
    Parameters are $N = 1000$, $k=2$, $d_{\mathrm{ave}} = 5$, $c_{\mathrm{out}} = 0.5$, and $\theta_i=1$ for SBM and $\theta_i \sim [\mathcal{U}(3,7)]^3$ for DCSBM.
    Simulations are averaged over $100$ runs.}
    \label{fig:robustness}
\end{figure}

\begin{table*}[t!]
 \centering
 \renewcommand{\arraystretch}{1.2}
 \caption{Detection performance on real social networks. Numbers in bold indicate the best performance and underlined ones are the second best.}
  \begin{tabular}{l|ccc|ccc|ccc|ccc}
   \hline
    & \multicolumn{3}{c|}{karate} & \multicolumn{3}{c|}{dolphins} & \multicolumn{3}{c|}{football} & \multicolumn{3}{c}{polblogs} \\
   Method&
   \multicolumn{1}{c}{$\varepsilon = 0.0$} & 
   \multicolumn{1}{c}{$0.1$}&
   \multicolumn{1}{c|}{$0.2$}&
   \multicolumn{1}{c}{$\varepsilon = 0.0$} &
   \multicolumn{1}{c}{$0.1$}&
   \multicolumn{1}{c|}{$0.2$}&
   \multicolumn{1}{c}{$\varepsilon = 0.0$} & 
   \multicolumn{1}{c}{$0.1$}&
   \multicolumn{1}{c|}{$0.2$}&
   \multicolumn{1}{c}{$\varepsilon = 0.0$} & 
   \multicolumn{1}{c}{$0.1$}&
   \multicolumn{1}{c}{$0.2$}\\
   \hline
    CIA & 
    0.8 & 0.58 & 0.54 & 0.97 & 0.78 & 0.71 & 0.78 & 0.69 & 0.55 & 0.75 & 0.7 & 0.63 \\ 
    SybilRank & 
    0.95 & 0.66 & \underline{0.56} & 0.96 & 0.76 & 0.61 & 0.82 & 0.73 & 0.59 & \underline{0.97} & \underline{0.96} & \underline{0.92} \\ 
    SybilWalk & 
    \underline{0.99} & 0.78 & \textbf{0.61} & \textbf{1.0} & \textbf{0.97} & \underline{0.77} & 0.88 & 0.77 & 0.59 & 0.75 & 0.77 & 0.75 \\ 
    SybilSCAR-C & 
    0.97 & 0.72 & 0.55 & \underline{0.99} & 0.82 & 0.7 & \underline{0.89} & 0.79 & 0.6 & \underline{0.97} & 0.93 & 0.84 \\ 
    SybilBelief & 
    \textbf{1.0} & \textbf{0.93} & 0.53 & \textbf{1.0} & \textbf{0.97} & \textbf{0.78} & \textbf{0.91} & \textbf{0.82} & \textbf{0.61} & \underline{0.97} & \underline{0.96} & 0.9 \\ 
    \textbf{Ours} & 
    \underline{0.99} & \underline{0.79} & 0.52 & \textbf{1.0} & \underline{0.93} & 0.74 & \underline{0.89} & \underline{0.8} & \underline{0.6} & \textbf{0.98} & \textbf{0.98} & \textbf{0.96} \\ 
   \hline
  \end{tabular}
  \label{tab:real_networks}
\end{table*}

\subsection{Robustness to label noise}
In practice, training datasets may contain noise due to human error~\cite{wang2013social}.
That is, some labeled benign (Sybil) nodes are actually Sybil (benign).
To evaluate the noise robustness of each method, we compare their detection performance when the node labels of training data with a fraction $\varepsilon ~(\le 0.5)$ are flipped.

Figure~\ref{fig:robustness} shows the detection performance of each method with respect to the fraction of label noise on synthetic graphs.
First, as discussed in Section~\ref{sect:5.1}, SybilBelief is quite robust against label noise because the corresponding filter can completely remove high frequency components, while CIA and SybilRank are not robust because of insufficient low-pass filtering.
SybilHeat has the second highest noise robustness after SybilBelief, and especially in the range for small label noise ($\varepsilon \le 0.1$), it is almost as robust as SybilBelief.
This is because the filter kernel $h(\lambda) = e^{-s\lambda}$ corresponding to SybilHeat greatly reduces the contribution of high frequency components.
Since the percentage of label noise in training datasets may not exceed $10\%$ in practice, we believe that SybilHeat performs well on real noisy datasets as well.
Although SybilWalk and SybilSCAR-C have the same filter kernel $h(\lambda)=1/\lambda$, SybilSCAR-C has lower noise robustness than SybilWalk.
This is because the contribution of high frequency components cannot be neglected since eigenvalues of the shift matrix corresponding to SybilSCAR-C on sparse graphs are aggregated around $\lambda=1$ (i.e., low and high frequency eigenvalues of $\Lmax$ are close to each other), as shown in Figure~\ref{fig:spec_dist}.

Table~\ref{tab:real_networks} shows the detection performance of each method for $\varepsilon=0.0$, $0.1$, and $0.2$ on real-world social networks.
As in the results on synthetic graphs, SybilBelief is robust to label noise on all datasets.
SybilWalk and SybilHeat are next to SybilBelief in robustness.
However, SybilHeat performs more consistently than SybilWalk because the performance of SybilWalk varies with the data, as shown in the results for polblogs.

\section{Conclusion}
We have shown that existing graph-based Sybil detection methods can be interpreted in a unified framework of low-pass filtering.
According to this interpretation, the performance of a Sybil detection method depends on how well low-pass filtering can extract informative low frequency components and remove noisy high frequency components.
In other words, for a Sybil detection method to perform well, 1) the filter kernel $h(\lambda)$ must properly emphasize (remove) low frequency (high frequency) components, and 2) the low frequency eigenvectors of the shift matrix $\bm{S}$ must have high community detectability.
Therefore, we have compared and analyzed existing detection methods from two perspectives (filter kernel properties and spectrum of the shift matrices) and have provided theoretical explanations of the superiority or inferiority of the performance and the conditions for high performance of each method.
Furthermore, we proposed the Sybil detection method (called SybilHeat) with the heat kernel as the filter kernel and the regularized Laplacian as the shift matrix, which satisfies the above two requirements.
SybilHeat is applicable to large social networks because it can be approximated by the Chebyshev polynomial approximation in the linear order with respect to the number of edges.
Numerical experiments show that SybilHeat performs consistently better than other methods on graphs with various structural properties.

Although we proposed a novel Sybil detection method using heat kernel and regularized Laplacian as the filter kernel and shift matrix, respectively, the performance might be improved by using other better filter kernels or shift matrices.
Also, as stated in existing studies~\cite{boshmaf2016integro, fu2017robust, gao2018sybilfuse, dorri2018socialbothunter}, learning node features and edge wights will improve detection performance.
We hope that this study leads to a deeper theoretical understanding and further improvement of graph-based Sybil detection methods.


%

\appendix[Linearization and Filtering Interpretation of Loopy Belief Propagation]
We first explain the linearization of loopy belief propagation by an approach presented in~\cite{mooij2005properties}.
We respectively define the node potential function and edge potential function as $\phi_i(s_i)=\exp(\beta \theta_i s_i)$ and $\psi_{ij}(s_i, s_j)=\exp(\beta J_{ij} s_i s_j)$ where $\beta$ is the inverse temperature $\theta_i$ is the local magnetic field on node $i$, and $J_{ij}$ is the interaction strength between node $i$ and node $j$.
Note that, the definitions of potential functions in~(\ref{eq:node_potential_function}) and~(\ref{eq:edge_potential_function}) can be recovered by normalizing the above definitions such that $\sum_{s_i}\phi_i(s_i)=1$ and $\sum_{s_{i}, s_j}\psi_{ij}(s_i, s_j)=1$.

Let us introduce the one-parametrized message $\nu_{ij} := \tanh^{-1}(\mu_{ij}(+1) - \mu_{ij}(-1) )$ instead of the message $\mu_{ij}(s_j)$.
For simplicity, denoting $\mu_{ij}^+ := \mu_{ij}(+1)$ and $\mu_{ij}^- := \mu_{ij}(-1)$, we obtain
\begin{align*}
    &\tanh(\nu_{ij}) \\
    &= \frac{1}{Z_{ij}} \left(e^{\beta J_{ij}} - e^{-\beta J_{ij}} \right) \left( e^{\beta \theta_i} \!\!\prod_{k \in \partial_{i \setminus j}}\!\! \mu_{ki}^+ - e^{-\beta \theta_i} \!\!\prod_{k \in \partial_{i \setminus j}}\!\! \mu_{ki}^- \right) \notag \\
    &= \tanh \left(\beta J_{ij} \right) ~
    \frac{\displaystyle e^{\beta \theta_i} \!\!\prod_{k \in \partial_{i \setminus j}}\!\! \mu_{ki}^+ - e^{-\beta \theta_i} \!\!\prod_{k \in \partial_{i \setminus j}}\!\! \mu_{ki}^- }{\displaystyle e^{\beta \theta_i} \!\!\prod_{k \in \partial_{i \setminus j}}\!\! \mu_{ki}^+ + e^{-\beta \theta_i} \!\!\prod_{k \in \partial_{i \setminus j}}\!\! \mu_{ki}^- }.
\end{align*}
Since
\begin{align*}
    \log \prod_{k \in \partial_{i \setminus j}}\!\! \left(\frac{ \mu_{ki}^+ }{\mu_{ki}^- } \right)^{\frac12}
    &= \sum_{k \in \partial_{i \setminus j}} \frac12 \log \frac{\mu_{ki}^+ + \mu_{ki}^- + \mu_{ki}^+ - \mu_{ki}^-}{ \mu_{ki}^+ + \mu_{ki}^- - \mu_{ki}^+ + \mu_{ki}^- } \\
    &= \sum_{k \in \partial_{i \setminus j}} \frac12  \log \frac{1 + (\mu_{ki}^+ - \mu_{ki}^-) }{ 1 - (\mu_{ki}^+ + \mu_{ki}^-) } \\
    &= \sum_{k \in \partial_{i \setminus j}} \tanh^{-1}(\mu_{ki}^+ - \mu_{ki}^-) 
    = \sum_{k \in \partial_{i \setminus j}} \!\!\nu_{ki},
\end{align*}
by using $\nu_{ij}$, we can rewrite~(\ref{eq:BP_message}) as
\begin{align}
    \tanh \!\left(\nu_{ij}^{\mathrm{new}} \right) = \tanh \bigl(\beta J_{ij} \bigr) \tanh \left(\beta \theta_i + \sum_{k \in \partial_{i \setminus j}}\!\! \nu_{ki} \right).
    \label{eq:nonlinear_operator}
\end{align}
The approximate marginal distribution $b_i(s_i)$ can also be one-parametrized by its expectation (called magnetization) $m_i = \langle s_i \rangle =  b_i(1) - b_i(-1)$ as follows:
\begin{align}
    m_i 
    &= \frac{\displaystyle e^{\beta \theta_i} \!\!\prod_{k \in \partial i} \mu_{ki}^\infty(1) - e^{-\beta \theta_i} \!\!\prod_{k \in \partial i} \mu_{ki}^\infty(-1) }{\displaystyle e^{\beta \theta_i} \!\!\prod_{k \in \partial i} \mu_{ki}^\infty(1) + e^{-\beta \theta_i} \!\!\prod_{k \in \partial i} \mu_{ki}^\infty(-1) } \notag\\
    &= \tanh \left( \beta \theta_i + \sum_{k \in \partial i} \nu_{ki}^\infty \right).
\end{align}

Denoting $\bm{\nu}=(\nu_{ij}) \in \mathbb{R}^{|\vec{E}|}$, let $\mathcal{BP}: \bm{\nu} \mapsto \bm{\nu}^{\mathrm{new}}$ be the nonlinear operator that maps $\bm{\nu}$ to $\bm{\nu}^{\mathrm{new}}$ by following~(\ref{eq:nonlinear_operator}).
The element of the Jacobian matrix $\bm{B} = \mathcal{BP}'(\bm{\nu}) \in \mathbb{R}^{|\vec{E}| \times |\vec{E}|}$ of $\mathcal{BP}$ is given by
\begin{align}
    B_{ij, kl} 
    = 
    \frac{\partial \nu_{ij}^{\mathrm{new}}}{\partial \nu_{kl}} 
    = \frac{\displaystyle \tanh(\beta J_{ij}) \!\left(1 - \tanh^2(h_{i \setminus j})\right)} {\displaystyle 1 -\tanh^2 (\beta J_{ij}) \tanh^2(h_{i \setminus j})}\, \delta_{il}(1 -\delta_{jk}),
    \label{eq:jacobian}
\end{align}
where $h_{i \setminus j} := \beta \theta_i + \sum_{k \in \partial_{i \setminus j}}\! \nu_{ki}$.
The matrix $\bm{B}$ is called a non-backtracking matrix and its $(ij, kl)$-element takes a non-zero value if two directed edges are consecutive (i.e., $i=l$) but do not back track (i.e., $j \neq k$).
To simplify the analysis, we assume the vanishing local field condition (i.e., $\theta_i=0$ for all $i \in V$).
This means that there is no prior information for each node.
In this case, we have $B_{ij, kl} = \tanh(\beta J_{ij}) \delta_{il}(1 -\delta_{jk})$ and thus~(\ref{eq:nonlinear_operator}) can be written as $\bm{\nu}^{\mathrm{new}} \approx \bm{B} \bm{\nu}$ by the linearization around the trivial fixed point $\bm{\nu}^*=0$.
Hence, when the spectral radius $\rho(\bm{B}) < 1$, the message $\bm{\nu}$ converges to the trivial fixed point $\bm{\nu}^*$.
On the other hand, when $\rho(\bm{B}) > 1$, $\bm{\nu}$ leaves from $\bm{\nu}^*$.
The eigenvector associated with an eigenvalue larger than $1$ is expected to correspond approximately to non-trivial (hopefully informative) fixed points of the loopy belief propagation~\cite{saade2016spectral}.

We consider unweighted non-backtracking matrix below (i.e., $B_{ij,kl} = \delta_{il}(1 -\delta_{jk})$).
For small $|\nu_{ki}|$, the magnetization of each node is given by
\begin{align}
    m_i = \tanh\left(\sum_{k \in \partial i}\nu_{ki}^\infty\right) \approx \sum_{k \in \partial i}\nu_{ki}^\infty.
    \label{eq:approximate_magnetization}
\end{align}
The eigenvector $\bm{\nu}$ associated with an eigenvalue $\eta > 1$ satisfies
\begin{align}
    \eta\nu_{ij} = \sum_{(k,l) \in \vec{E}} \!\! B_{ij, kl}\nu_{kl} = \sum_{k \in \partial_{i \setminus j}} \!\!\nu_{ki} = m_i - \nu_{ji}.
\end{align}
Similarly, $\eta \nu_{ji} = m_j - \nu_{ij}$ is hold.
Thus, we have $\nu_{ij} = (\eta m_i - m_j)/(\eta^2 - 1)$.
By substituting this into~(\ref{eq:approximate_magnetization}), we obtain
\begin{align}
    (\eta^2 - 1)m_i + |\partial i| m_i - \eta \sum_{k \in \partial i} m_k = 0.
\end{align}
Alternatively, we rewrite the above equation as $\bm{H}(\eta)\bm{m}=0$ by using the Bethe-Hessian matrix.

The small eigenvalues of $\bm{H}(r)$ are closely related to the informative eigenvalues of $\bm{B}$, and the corresponding eigenvectors approximately give the magnetization $\bm{m}$ calculated by the loopy belief propagation~\cite{saade2014spectral, saade2016spectral}.
On the basis of this observation, a community detection algorithm using eigenvectors corresponding to small (low frequency) eigenvalues of the Bethe-Hessian matrix has been proposed and its performance has been demonstrated to be comparable to loopy belief propagation~\cite{saade2014spectral, dall2019revisiting, dall2020optimal}.
In the same spirit, we can interpret SybilBelief as low-pass filtering as in~(\ref{eq:SybilBelief_lowpass}) using the ideal low-pass filter kernel $g(\omega)$ that extracts only the low frequency spectrum of $\bm{H}(r)$.

Note that we have assumed the vanishing local field condition through our analysis.
This assumption is quite strong since it implies ignoring all known node labels.
However, since the local magnetic field helps accelerate the convergence of the message and biases it toward the desired local minimum~\cite{knoll2017loopy}, a similar effect is expected by low-pass filtering of the known label vector $\bm{q}$.



\ifCLASSOPTIONcaptionsoff
  \newpage
\fi



%

\bibliographystyle{IEEEtran}
\bibliography{IEEEabrv,ref}


%

\begin{IEEEbiography}[{\includegraphics[width=1in,height=1.25in,clip,keepaspectratio]{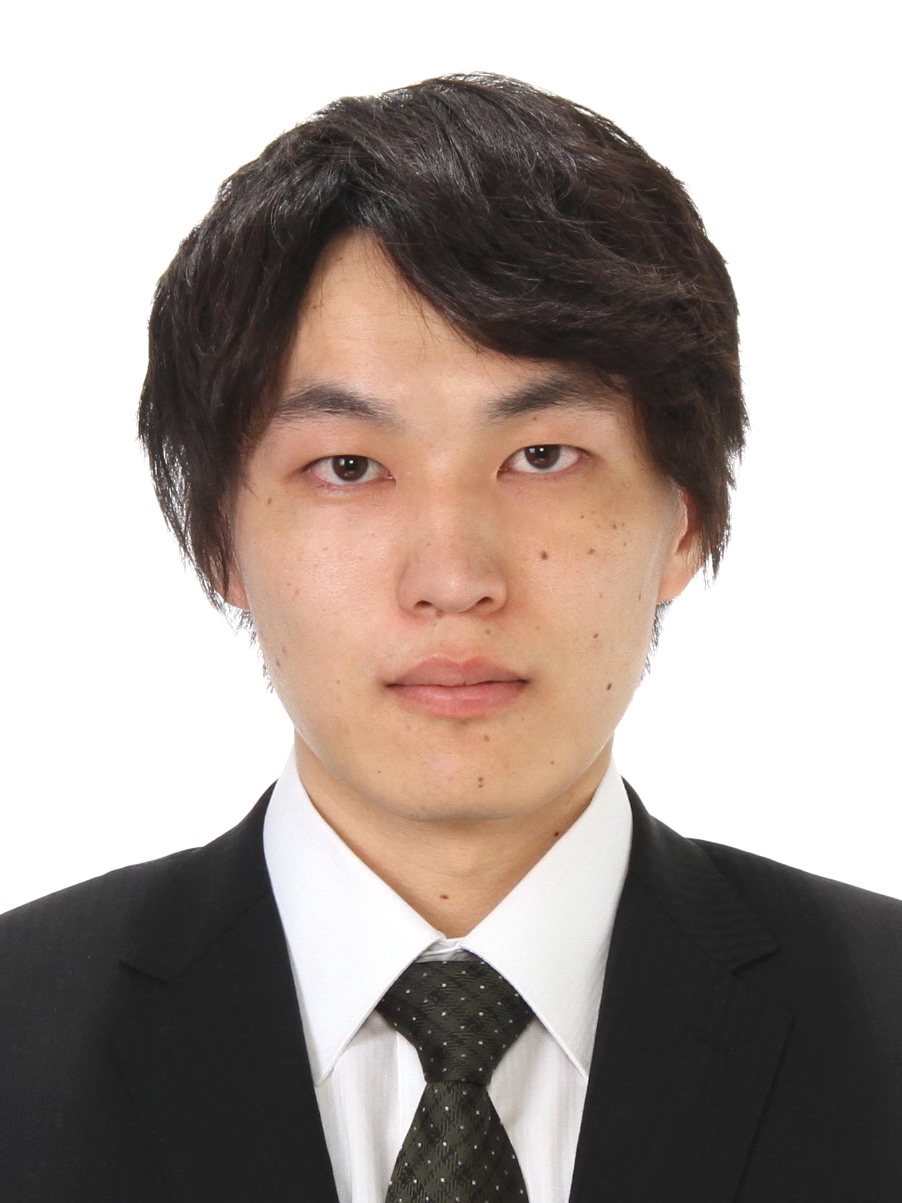}}]{Satoshi Furutani}
received his B.E. and M.E. degrees in System Design Engineering from Tokyo Metropolitan University, Japan, in 2016 and 2018, respectively.
Since joining Nippon Telegraph and Telephone Corporation (NTT) in 2018, he has been engaged in research and development on network science. 
He is currently a researcher at NTT Social Informatics Laboratories and is pursuing a Ph.D. degree at Tokyo Metropolitan University.
His research interests include analysis of social network dynamics and graph signal processing.
He is a member of IPSJ and IEICE.
\end{IEEEbiography}

\begin{IEEEbiography}[{\includegraphics[width=1in,height=1.25in,clip,keepaspectratio]{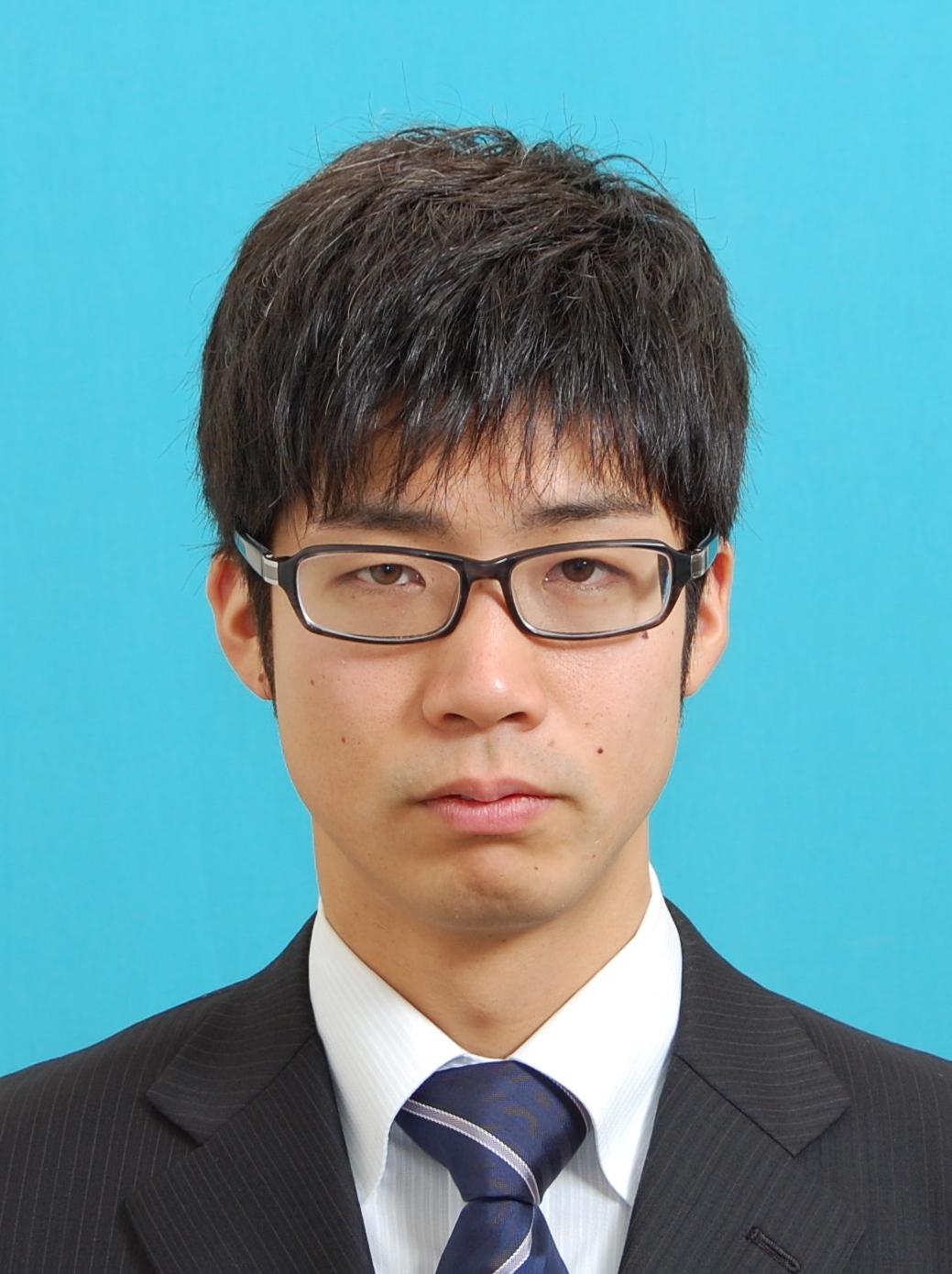}}]{Toshiki Shibahara}
is currently a researcher at NTT Social Informatics Laboratories, Tokyo, Japan.
He received his B.E. degree in engineering and M.E degree in information science and technology from The University of Tokyo, Japan in 2012 and 2014.
He also received his Ph.D. degree in information science and technology from Osaka University, Osaka, Japan in 2020.
Since joining Nippon Telegraph and Telephone Corporation (NTT) in 2014, he has been engaged in research on cyber security and machine learning.
\end{IEEEbiography}

\begin{IEEEbiography}[{\includegraphics[width=1in,height=1.25in,clip,keepaspectratio]{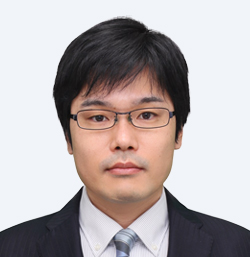}}]{Mitsuaki Akiyama}
received his M.E. and Ph.D. degrees in information science from Nara Institute of Science and Technology in 2007 and 2013. Since joining Nippon Telegraph and Telephone Corporation (NTT) in 2007, he has been engaged in research and development on cybersecurity. He is currently a Senior Distinguished Researcher at NTT Social Informatics Laboratories. He received Cybersecurity Encouragement Award of the Minister for Internal Affairs and Communications in 2020, ISOC NDSS 2020 Distinguished Paper Award in 2020, and IPSJ/IEEE Computer Society Young Computer Researcher Award in 2022. His research interests include cybersecurity measurement, offensive security, and usable security and privacy. He is a senior member of IPSJ and a member of IEEE and IEICE. 
\end{IEEEbiography}

\begin{IEEEbiography}[{\includegraphics[width=1in,height=1.25in,clip,keepaspectratio]{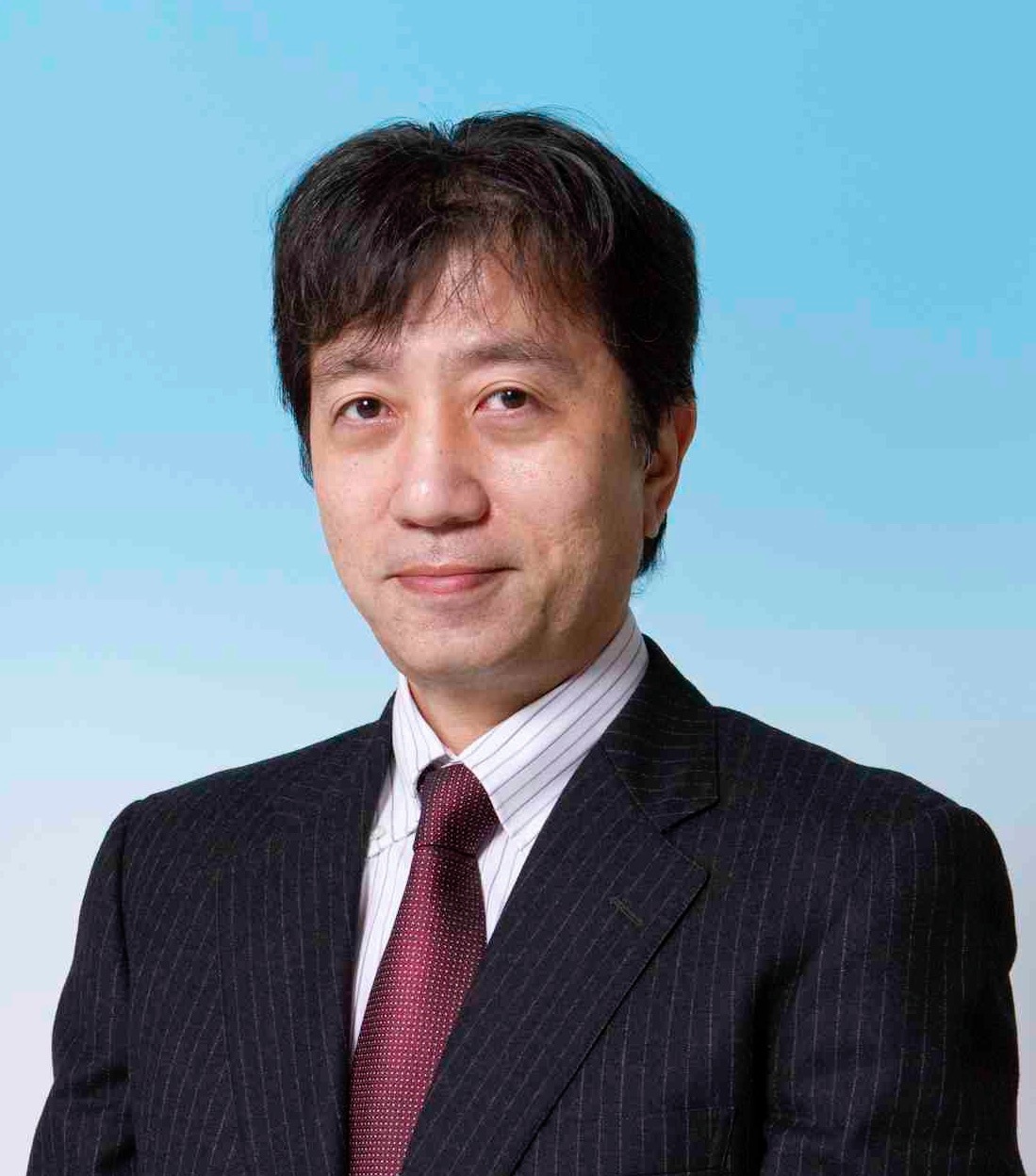}}]{Masaki Aida} 
received his B.S. degree in Physics and 
M.S. degree in Atomic Physics from St.~Paul's University, Tokyo, Japan, in 1987 and 1989, respectively, and his Ph.D. in Telecommunications Engineering from the University of Tokyo, Japan, in 1999. 
In April 1989, he joined NTT Laboratories. 
From April 2005 to March 2007, he was an Associate Professor at the Faculty of Systems Design, Tokyo Metropolitan University. 
He has been a Professor of the Graduate School of Systems Design, Tokyo Metropolitan University since April 2007. 
His current interests include analysis of social network dynamics and distributed control of computer communication networks. 
He received the Best Tutorial Paper Award and the Best Paper Award of IEICE Communications Society 
in 2013 and 2016, respectively, and IEICE 100-Year Memorial Paper Award in 2017. 
He is a fellow of IEICE, a senior member of IEEE, and a member of ACM and ORSJ.
\end{IEEEbiography}





\end{document}